\documentclass{nature}

\usepackage[dvips]{graphicx}

\title{A massive star origin for an unusual helium-rich supernova
in an elliptical galaxy}

\author{K. S. Kawabata$^{1}$,
K. Maeda$^{2}$, K. Nomoto$^{2}$,
S. Taubenberger$^{3}$, M. Tanaka$^{4,2}$,
J. Deng$^{5}$, E. Pian$^{6}$, T. Hattori$^{7}$ 
\& K. Itagaki$^{8}$}

\begin{document}

\maketitle

{\footnotesize\it Accepted by Nature (24 March 2010)}

\begin{affiliations}
 {\footnotesize
 \item Hiroshima Astrophysical Science Center, Hiroshima University, 
 1-3-1 Kagamiyama, Higashi-Hiroshima, Hiroshima 739-8526, Japan.
 \item Institute for the Physics and Mathematics of the Universe (IPMU), 
 University of Tokyo, 5-1-5 Kashiwanoha, Kashiwa, Chiba 277-8583, Japan.
 \item Max-Planck-Institut f\"ur Astrophysik, 
 Karl-Schwarzschild-Stra{\ss}e 1, 85741 Garching, Germany.
 \item Department of Astronomy, School of Science, University of Tokyo, 
 Bunkyo-ku, Tokyo 113-0033, Japan.
 \item National Astronomical Observatories, CAS, 20A Datun Road,
 Chaoyang District, Beijing 100012, China.
 \item INAF Osservatorio Astronomico di Trieste, Via Tiepolo 11, I-3413
 Trieste, Italy. 
 \item Subaru Telescope, National Astronomical Observatory of Japan, 
 Hilo, HI 96720.
 \item Itagaki Astronomical Observatory, Teppo-cho, Yamagata 990-2492, Japan.
 }
\end{affiliations}

\begin{abstract}

The unusual helium-rich (type Ib) supernova SN 2005E is distinguished
from any supernova hitherto observed by its faint and rapidly fading
light curve, prominent calcium lines in late-phase spectra and lack
of any mark of recent star formation near the supernova location. 
These properties are claimed to be explained by a helium detonation in
a thin surface layer of an accreting white dwarf (ref. 1).  Here we
report on observations of SN 2005cz appeared in an elliptical
galaxy, whose observed properties resemble those of SN 2005E 
in that it is helium-rich and unusually faint, fades rapidly, 
shows much weaker oxygen emission lines than those of calcium in 
the well-evolved spectrum. We argue that these properties are 
best explained by a core-collapse supernova at the low-mass end 
($8-12 M_{\odot}$) of the range of massive stars that explode (ref. 2).
Such a low mass progenitor had lost its hydrogen-rich envelope 
through binary interaction, having very thin oxygen-rich and 
silicon-rich layers above the collapsing core, thus ejecting 
a very small amount of radioactive $^{56}$Ni and oxygen.
Although the host galaxy NGC 4589 is an elliptical, some studies
have revealed evidence of recent star-formation activity (ref. 3),
consistent with the core-collapse scenario.

\end{abstract}

We discovered SN 2005cz on 2005 July 17.5 UT in the 
elliptical galaxy NGC 4589.
The spectrum of SN 2005cz taken on July 28 is 
well consistent with post-maximum spectra of type Ib supernovae 
(SNe Ib)$^{4}$.
Thus, SN 2005cz would originate from a core-collapse of an 
envelope-stripped massive star.
We tentatively assume that the epoch of our first spectrum
is at $t=+26$ days, where $t$ is time after the maximum brightness 
(Fig. 1; see SI \S 1).

The late-time spectrum of SN 2005cz at $t=+179$ days is very unique; 
unlike most of other SNe Ibc/IIb SN 2005cz shows much stronger 
[Ca II] $\lambda\lambda$7291, 7323 than [O I] $\lambda\lambda$6300, 6364 
(Fig. 2; see ref. 12, 13 for other SNe with large Ca/O and
SI \S3 for comparative discussion.). 
Oxygen is ejected mostly from the oxygen layer formed during
the hydrostatic burning phase; its mass depends sensitively on the
progenitor mass and is smaller for lower-mass progenitors.
On the other hand, Ca is explosively synthesized during the explosion.
Theoretical models predict that the stars having main-sequence masses of 
$M_{\rm ms}=13M_{\odot}$ and $18 M_{\odot}$ produce $0.2$ and $0.8 M_{\odot}$
of O, and $0.005$ and $0.004 M_{\odot}$ of Ca, respectively (e.g., ref. 14).
Therefore, the Ca/O ratio in the SN ejecta is sensitive to the 
progenitor mass$^{15,16}$.
To produce the extremely large Ca/O ratio, the mass of the 
progenitor star of SN 2005cz should be smaller than of any other SNe Ib 
reported to date.
For both SNe 1993J and 1994I that show weaker [Ca II] than [O I] 
(Fig. 2), the progenitor's masses are estimated to be 
$\sim 12-15 M_{\odot}$ (ref. 17,18),
which are the smallest among well-studied samples
with [Ca II]$<$[O I] (e.g., ref. 16,19,20; see also Supplementary Fig. 3).
Thus, the progenitor mass of SN 2005cz is likely $\leq 12-15 M_{\odot}$.

SN 2005cz is intrinsically fainter than the well-studied SN Ic 1994I
by $\Delta R\sim 1.5$ mag (Fig. 3).
In the pseudo bolometric light curve, the decline rate 
from the intermediate to the late phase is consistent with 
$(M_{{\rm ej},\odot}^2/E_{51}) \leq 1$, and the luminosity 
requires that $M(^{56}\rm Ni)\leq 0.005-0.02 M_{\odot}$ (Fig. 4).
Additionally, $(M_{{\rm ej},\odot}/E_{51}) \sim 1$ is suggested 
from the line velocity (Fig. 4 legend).
We thus estimate $M_{{\rm ej}, \odot} \leq 1$ and $E_{51} \leq 1$, 
indicating a small progenitor mass 
($\leq 12 M_{\odot}$; ref. 27, 2). 

To explain the above peculiarities, we suggest a star with 
$M_{\rm ms} = 10-12 M_{\odot}$ as the most likely origin 
of SN 2005cz.
If such a star had been single, its mass (and thus its mass loss rate)
would have been too small to lose most of its H-rich envelope. 
Thus this star must have been in a close binary system.
Then it became a He star of $\sim 2.5 M_{\odot}$ 
after undergoing Roche lobe overflow.
This He star formed a C+O core of $\sim 1.5 M_{\odot}$, 
which marginally exceeded the lower mass limit to form a Fe core$^{28,29}$.  
The overlying He layer had $\sim 1 M_{\odot}$.
Eventually, the He star underwent Fe core-collapse to explode as 
a SN Ib, leaving a $\sim 1.5 M_{\odot}$ neutron star behind.  
The ejecta had $\sim 1 M_{\odot}$, consistent with the 
observed constraint.
The ejecta consists mostly of unburned material in the He layer
and a small amount of explosively synthesized elements.
The explosive burning products contain some Fe, Ca, S, and Si, 
but not much oxygen.
Also the ejected part of the unburned oxygen-rich layer is extremely small.
This scenario can explain the peculiar nebular spectrum with large 
[Ca II]/[O I] ratio, as well as the low luminosity and its relatively 
rapid decrease.

An alternative candidate of the progenitor is a star with 
$M_{\rm ms} \sim 8-10 M_{\odot}$ in a close binary system.
Such a star forms an electron-degenerate ONeMg core and 
undergo electron-capture-induced collapse$^{29}$.
The most likely scenario to realize a SN Ib would be the
merging of an ONeMg white dwarf and a He white dwarf.
The delay time between the star formation and the merging 
could be long enough to explain the origin of both 
SN 2005cz and recently reported 2005E$^1$ with this scenario.

As for the host galaxy problem, the $\sim 10M_{\odot}$ star model
is found to be consistent with the properties recently-inferred for 
the host galaxy of SN 2005cz.
It is still a genuine E2 galaxy$^{30}$, but has a relatively
young stellar population with life times of $\sim 10^7-10^8$ 
years$^{3}$ and SN Ib 2005cz is likely the end product of one of 
these young stars (See SI \S2).

The mass range of 10-12 $M_{\odot}$ has not been 
theoretically investigated in much detail so far,
but, as SN 2005cz suggests, the SNe 
resulting from these stars may have a very special abundance 
pattern in the ejecta and play an important role in the chemical
evolution of galaxies (see SI \S3).

\begin{addendum}
 \item[Acknowledgements]
 We would like to thank D. Leonard for permission to publish his 
 early-time spectrum of SN 2005cz.
 We also thank S. Valenti for permission to use their spectra
 of SN 2008ha.
 We gratefully acknowledge advice and help from P. A. Mazzali
 through a series of this study. 
 This work is based on observations collected at the 
 2.2-m Telescope at the Calar Alto Observatory (Sierra de Los Filabres,
 Spain), at the Keck Telescope, and at Subaru Telescope (operated 
 by the National Astronomical Observatory of Japan).
 We are grateful to the staff members at the observatories 
 for their excellent assistance, especially to T. Sasaki,
 K. Aoki, G. Kosugi, T. Takata and M. Iye. 
 This research is supported by World Premier International Research Center
 Initiative (WPI Initiative), MEXT, Japan, and by the Grant-in-Aid for
 Scientific Research of the JSPS and MEXT.
 M.T. is supported through the JSPS (Japan Society for the 
 Promotion of Science) Research Fellowship for Young Scientists.
 J.D. is supported by the NSFC and by the 973 Program of China.
 \item[Author Contributions] K.S.K., K.M., K.N., J.D.
 and E.P. organized the observations and discussion. 
 K.M., K.N and K.S.K. have written the manuscript.
 K.S.K., S.T., and K.I. are responsible for data acquisition 
 and reduction; J.D. and E.P. were the PIs of the relevant Subaru 
 programs, S05B-132 and S05B-054, respectively.
 M.T., S.T. contributed to discussions. 
 T.H. provided expertise on data acquisition at Subaru Telescope.
 \item[Competing Interests] The authors declare that they have no
competing financial interests.
 \item[Correspondence] Correspondence and requests for materials
should be addressed to K.S.K. (email:\\ kawabtkj@hiroshima-u.ac.jp).
\end{addendum}

 \begin{center}
  \includegraphics[width=0.8\textwidth]{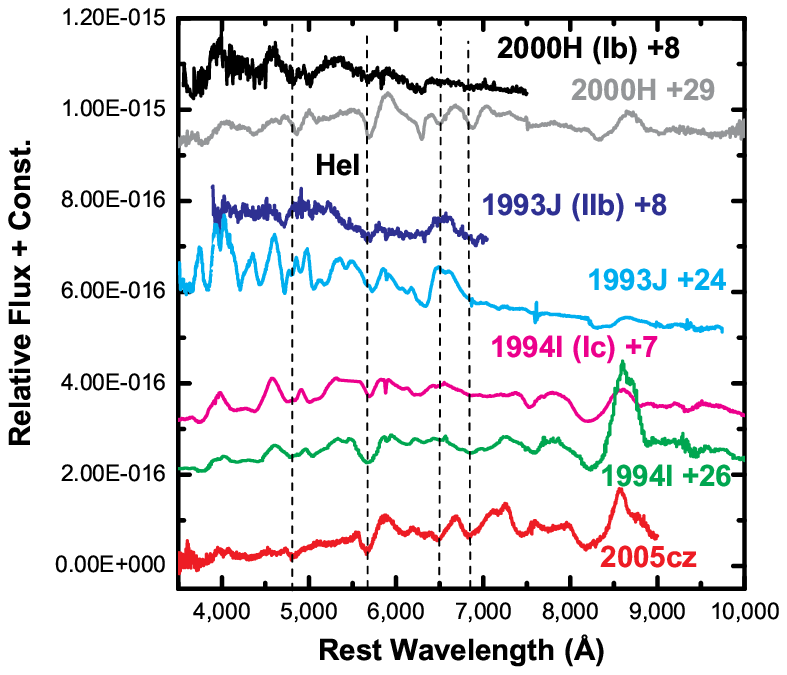}
  \includegraphics[width=0.8\textwidth]{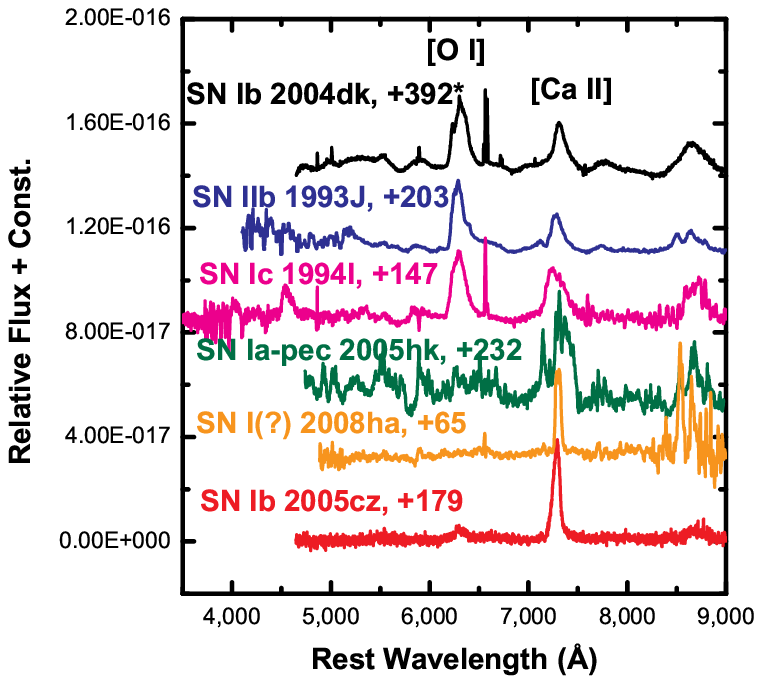}
  \includegraphics[width=0.8\textwidth]{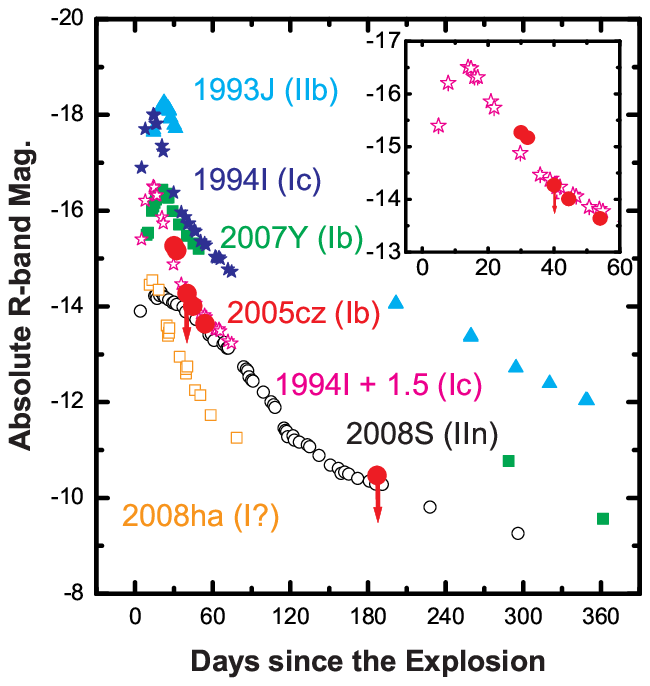}
  \includegraphics[width=0.8\textwidth]{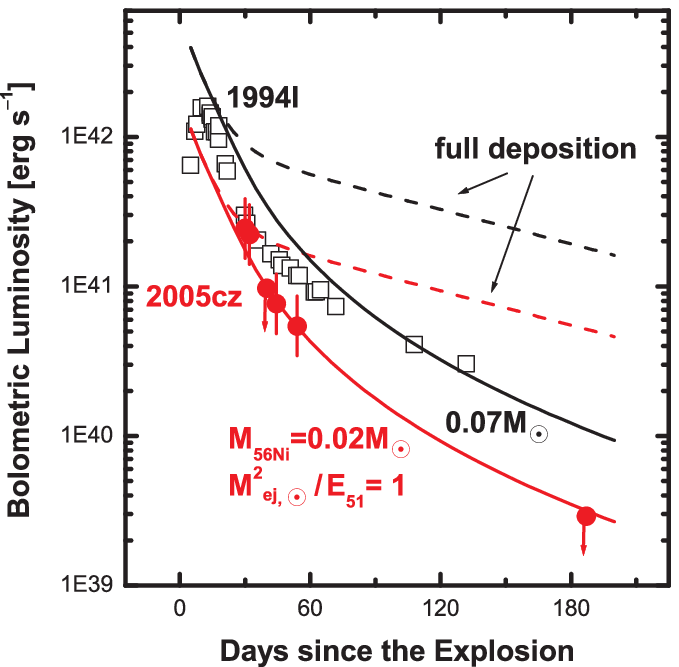}
 \end{center}

\clearpage

\noindent
\sffamily\noindent\textbf {Figure 1}\hspace{2em}
{\sf Early-time spectrum of SN Ib 2005cz compared with other
 envelope-stripped SNe at similar phases. 
 The spectrum of SN 2005cz (red) is taken on 2005 Jul 28. 
 Also shown are the spectra of SN Ib 2000H at $t=+8$ (black) and 
 $t=+29$ days (gray)$^{5}$, 
 SN IIb 1993J at $t=+8$ days (blue) and 
 $t=+24$ days (cyan)$^{6}$,  and 
 SN Ic 1994I at $t=+7$ days (magenta) and $t=+26$ days (green)$^{7,8}$.
 The SN Ib is characterized by strong helium lines and weak silicon 
 lines, while in the SN Ic both helium and silicon lines are weak.
 The SN IIb shows a SN II-like spectrum characterized 
 by the strong hydrogen features at early times, and becomes 
 SN Ib/c-like at late times.
 All these SNe are thought to have partly or fully stripped off 
 their outer layers of hydrogen and helium before the explosions.
 The overall appearance of spectral features in SN 2005cz is 
 quite similar to those of 
 the SN Ib 2000H at $t=+29$ days, the SN IIb 1993J 
 at $t=+24$ days (despite its stronger H lines), and also 
 the typical SN Ic 1994I at $t=+26$ days (despite its lack 
 of the strong He lines).
 The spectra are corrected for the host redshift and the reddening.
 We adopted a total (Milky Way + host) reddening of 
 $E(B-V)=0.13$ $(0.03+0.1)$ mag in SN 2005cz, 
 $0.23$ $(0.23+0.0)$ mag in SN 2000H, 
 $0.45$ mag in SN 1994I, and
 $0.3$ mag in SN 1993J.
 The flux is on an absolute scale for SN 2005cz, calibrated with the 
 Calar Alto photometry obtained four nights later. 
 For the comparison SNe, the fluxes are on an arbitrary scale 
 and constants are added for presentation. 
 The positions of the prominent He I lines 
 are shown by the dotted lines.
 The spectrum of SN 2005cz is well consistent with the post-maximum 
 spectra of SNe Ib.
}

\noindent
\sffamily\noindent\textbf {Figure 2}\hspace{2em}
{\sf Calcium-rich late-time spectrum of SN Ib 2005cz.
 It is taken on 2006 Dec 27 ($t=+179$ days). 
 Also shown are SN Ib 2004dk at $t\sim 390$ days$^{9}$,
 SN IIb 1993J at $t=+203$ days$^{6}$, 
 SN Ic 1994I at $t=+147$ days$^{8}$, 
 peculiar SN Ia 2005hk at $t=+232$ days$^{10}$,
 and peculiar SN I(?) 2008ha at $t=+65$ days$^{11}$.
 As time goes by, the ejecta become transparent to optical light, 
 following the expansion and density decrease. 
 Late-time spectra of SNe Ib/c are thus characterized by various 
 emission lines, mostly of forbidden transitions. 
 The spectrum of SN 2004dk is typical for SNe Ib/c at late times
 (e.g., see fig.2 of ref. 9).
 The spectra are corrected for the host redshift, but not for reddening. 
 The flux is on an approximate absolute scale for SN 2005cz, 
 calibrated with the spectroscopic standard star 
 (but not with photometry), while it is on an 
 arbitrary scale for the comparison SNe.
 The asterisk of SN 2004dk denotes the days since its discovery 
 (not maximum light).
 It is very unique that SN 2005cz shows only weak [O I] 
 $\lambda\lambda$6300, 6364 and much stronger 
 [Ca II] $\lambda\lambda$7291, 7323 than [O I].
 The relatively weak Ca II IR triplet compared with other SNe
 might suggest a lower density ejecta of SN 2005cz.
 It is interesting that the [Ca II] line is considerably narrow 
 (half-width at half-maximum $0.005c$) compared with the 
 blueshift of the absorption in Ca II IR triplet in the 
 early-time spectrum ($\sim 0.04c$).
}

\noindent
\sffamily\noindent\textbf {Figure 3}\hspace{2em}
{\sf Absolute $R$-band light curve of rapidly-fading SN Ib 2005cz.
 It is shown by filled red circles and compared with those of 
 SN IIb 1993J (cyan triangles), 
 SN Ic 1994I (blue stars), SN Ib 2007Y (green squares), 
 SN IIn 2008S (black open circles), and SN I (?) 2008ha (orange open squares). 
 Also shown is the light curve of SN 1994I, but dimmed by 1.5 magnitudes 
 (magenta open stars).
 For SN 2005cz, the first three points denote unfiltered 
 magnitudes which are approximately $R$-band magnitudes. 
 The two points with downward arrows are 
 $3\sigma$ upper-limits.
 The distance moduli and total reddening values are taken as follows: 
 [$\mu$, E(B-V)] = (32.23 mag, 0.13 mag) for 2005cz (see SI \S1), 
 (27.8 mag, 0.3 mag) for 1993J, 
 (29.6 mag, 0.45 mag) for 1994I, 
 (31.43 mag, 0.112 mag) for 2007Y, 
 (31.55 mag, 0.076 mag) for 2008ha, and 
 (28.78 mag, 0.687 mag) for 2008S. 
 We assume $R_{V} = 3.1$ to convert the colour excess 
 to the $R$-band extinction. 
 The data points, as well as the distance and the reddening, 
 are from the literature $^{6,11,21,22,23}$.
 The putative explosion date for SN 2005cz is assumed to be 2005 Jun 17, 
 30 days before the discovery and 15 days before maximum brightness
 (SI \S1).
 The LC tail of SN 2005cz is similar to those of SN IIn 2008S 
 and SN Ic 1994I (dimmed by $1.5$ mag).
 From this, we estimate the mass of $^{56}$Ni as 
 $M(^{56}\rm Ni) = 10^{-1.5/2.5} \times 0.07 M_{\odot} \sim 0.018 M_{\odot}$
 ($M(^{56}\rm Ni)=0.07 M_{\odot}$ is for SN 1994I$^{24}$).
}

\noindent
\sffamily\noindent\textbf {Figure 4}\hspace{2em}
{\sf Pseudo bolometric light curve of SN Ib 2005cz
 suggests that the ejecta has a low mass, low kinetic
 energy, and a tiny amount of $^{56}$Ni.
 The light curve (filled red circles) is compared with 
 a simple $\gamma$-ray and positron deposition model with
 $M(^{56}\rm Ni) = 0.02 M_{\odot}$ 
 and $M_{{\rm ej}, \odot}^2/E_{51} = 1$ (red line),
 where $E_{51}$ is the kinetic energy $E_{\rm K}$ measured in 
 unit of $10^{51}$ ergs.
 We also plot the bolometric light curve of SN Ic 1994I 
 (open black squares)$^{21}$ and a simple deposition model with 
 $M(^{56}\rm Ni) = 0.07 M_{\odot}$ (black line) for comparison.
 Except for the last point (upper-limit), we simply assume the 
 bolometric correction $BC \equiv M_{\rm Bol} - M_{R} = 0.5$, 
 derived from SNe 1998bw, 2002ap and 2008D at similar epochs
 $^{25,26,27}$.
 As this is a very crude estimate, we adopt an error bar of 
 $\pm 0.5$ mag for the bolometric luminosity.
 The deposition models adopt the $\gamma$-ray opacity for the
 Compton scattering ($\tau_{\gamma} \propto 
 M_{\rm ej}^2 {E_{\rm K}}^{-1} t^{-2}$)
 and assuming the full deposition of positrons.
 The decline rate from the intermediate to the late phase is 
 consistent with $(M_{{\rm ej}, \odot}^2/E_{51}) \leq 1$.
 Combining this expression with ($M_{{\rm ej}, \odot}/E_{51}) \sim 1$
 as indicated by the similarity in the absorption velocity 
 seen in SN 2005cz and those in SNe 1993J and 1994I (Fig. 1, 
 Supplementary Fig. 1), we estimate  $M_{{\rm ej}, \odot} \leq 1$ 
 and $E_{51} \leq 1$. The luminosity requires that 
 $M(^{56}\rm Ni) \leq 0.02 M_{\odot}$. Note that the estimate 
 for $M(^{56}\rm Ni)$ is sensitively affected by the explosion 
 date. The upper limit to $M(^{56}\rm Ni)$ is only 
 $M(^{56}\rm Ni) \leq 0.005 M_{\odot}$, if the explosion date
 is as late as 2005 Jul 15  (just a few days before the discovery).
 }

\clearpage
\noindent
{\LARGE Supplementary Information}

\section{Observation}

We discovered SN 2005cz on 2005 July 17.5 UT at $13''$ offset 
from the nucleus of the elliptical galaxy NGC 4589.
No object brighter than 18.5 mag was visible at the SN 
position on 2005 June 20 (ref. 31).

The discovery and subsequent unfiltered images were 
taken with a 0.6-m reflector and a CCD (Kodak KAF-1001E) 
at Itagaki Astronomical Observatory (IAO) in Yamagata, Japan.
The derived magnitudes are approximately consistent with $R$ 
magnitudes.

Other imaging observations were performed by the Calar Alto 
2.2-m telescope (CA2.2) equipped with the
Calar Alto Faint Object Spectrograph (CAFOS)
in $B$, $V$, $R$ and $I$ bands, and by the 8.2-m Subaru Telescope 
equipped with the Faint Object Camera and Spectrograph 
(FOCAS$^{32}$) in $B$ and $R$ bands. 
Imaging observations with Subaru were done in photometric 
conditions; standard stars around PG 1525-071 in August 2005
and around PG 0942-029 in December 2005 were observed for 
photometric calibration.

The data reduction was performed using the IRAF 
package DAOPHOT. 
Since SN 2005cz was close to the bright core of the host galaxy,
we subtracted the host galaxy component prior to the photometry
for more reliable photometry. 
For the unfiltered images at IAO, we used the pilot survey image 
taken with the same system on 2005 May 25 
as the host template.\footnote{The discovery magnitude of 16.0 reported in 
IAUC 8569 included a large bias caused by the steep brightness
distribution of the host galaxy core.}
For the CA2.2 and Subaru images, we took data of the same field,
as well as the standard stars around PG 0918+029,
by CA2.2+CAFOS in a photometric night on 2009 Feb 19 and
used them as the host galaxy template after point-spread-function-matching.

The derived $R$ (and unfiltered) magnitudes are shown in 
the Supplementary Table 1. 
The other magnitudes are $B = 21.18 \pm 0.30$ mag, 
$V = 19.69 \pm 0.18$ mag, $I = 17.72 \pm 0.11$ mag on 2005 Aug 1, and
$B = 21.0 \pm 0.2$ mag on 2005 Aug 10.
On 2005 Jul 27 and Dec 27, the SN was not detected and we just derived 
$3 \sigma$ upper-limit for the SN luminosity.
For the Dec 27 data, we derived more reliable upper-limit
of the bolometric luminosity (Fig. 4) as follows; 
first, we subtracted a continuum from the Subaru spectrum and
then scaled it to the observed upper-limit ($R > 22.1$). 
We then integrated the flux at optical wavelengths. 
Finally, we assumed that the NIR contribution was 30\% of 
the optical luminosity, a typical value seen in SNe Ib/c 
at late phases$^{33}$.

The early-time spectrum of SN 2005cz was obtained on 2005 Jul 28 UT
with the 10-m Keck I Telescope equipped with the
Low-Resolution Imaging Spectrometer (LRIS$^{34}$).
The total exposure time was 500 s.
The seeing was $\sim 1''.2$ and the airmass was relatively large, $\sim 2.3$. 
The wavelength resolution measured from sky lines was 9 \AA.
The late-phase spectrum was obtained with Subaru+FOCAS on 2005 Dec 27 UT.
The total exposure time was 1800 s.
The seeing was $\sim 1''.0$ and the airmass was $\sim 1.8$. 
The wavelength resolution was 11 \AA.
These data were reduced with IRAF in a standard manner for long-slit
spectroscopy.

For the extinction within our Galaxy and the host galaxy,
we adopted $E(B-V)=0.03$ and $0.1$ mag, respectively.
The former is inferred from the infrared dust map$^{35}$, 
while the latter is estimated from the equivalent width
of Na I D absorption feature in the early-phase spectrum ($EW \leq 0.34$\AA )
and an empirical formula$^{36}$.
Although the formula allows for a range of 
$E(B-V) \leq 0.044 - 0.13$ mag, the extinction within the inner $3''$ of 
NGC 4589 has been estimated to be $E(B-V) \sim 0.1$ mag by the 
spectrum template fitting$^{37}$. 
Since the SN position 
is reasonably separated from the dusty bar near the nucleus 
of the host galaxy, the extinction should not be large there. 
Therefore, we take $E(B-V) = 0.1$ mag as a reference value 
for the host extinction. 

We assume that $t=+26$ days as the epoch of the first spectrum 
(Jul 28) because of the overall resemblance of 
the spectral features with SN Ic 1994I at $t=+26$ days and with
SN Ib 2000H at $t=+29$ days.
This estimate still includes a large uncertainty; e.g., the He I 
line velocities of $\sim 9,000 - 12,000$ km s$^{-1}$ on Jul 28 
(Supplementary Fig. 1) would be more typical for SNe Ib at 
$t=0$ to $+10$ days.
Since the estimation for $^{56}$Ni mass is 
sensitively affected by the explosion date, we also consider
an extreme case in which the discovery was close to the 
explosion date, and give the possible range of $M(^{56}\rm Ni)$
(see Fig. 4 legend).
Anyway, the choice here does not affect our main conclusions.

\clearpage
\begin{table*}
{\footnotesize
\begin{center}
{Supplementary Table 1: Summary of observation of SN 2005cz}
\begin{tabular}{lllllc}
\hline
\hline
Date           & MJD     & Epoch    & Telescope+   & Setup                      & $R$   \\
(UT)           &         & (days)   & Instrument   &                            & (mag) \\
\hline
2005 May 25.7  & 53516.7 &  ---     & IAO 0.6      & Imaging(unfiltered)        & --- \\
2005 Jul 17.5  & 53568.5 &  $+15.5$ & IAO 0.6      & Imaging(unfiltered)        & $17.3\pm 0.1$ \\
2005 Jul 19.5  & 53570.5 &  $+17.5$ & IAO 0.6      & Imaging(unfiltered)        & $17.4\pm 0.1$ \\
2005 Jul 27.5  & 53578.5 &  $+25.5$ & IAO 0.6      & Imaging(unfiltered)        & $>18.3$ \\
2005 Jul 28    & 53579   &  $+26$   & KeckI+LRIS   & Spectropolarimetry         & --- \\
2005 Aug  1.0  & 53583.0 &  $+30.0$ & CA2.2+CAFOS  & Imaging($BVRI$)            & $18.56\pm 0.12$ \\
2005 Aug 10.3  & 53592.3 &  $+39.3$ & Subaru+FOCAS & Imaging($BR$)              & $18.93\pm 0.05$ \\
2005 Dec 27.6  & 53731.6 & $+178.6$ & Subaru+FOCAS & Imaging($BR$)/Spectroscopy & $>22.1$ \\
2009 Feb 19.0  & 54881.0 &  ---     & CA2.2+CAFOS  & Imaging($BVRI$)            & --- \\
\hline         
\end{tabular}
\end{center}
}
\end{table*}

\section{Stellar population in the elliptical galaxy NGC 4589}

It is apparently puzzling that SN Ib 2005cz appears in the 
elliptical galaxy NGC 4589 if it is the Fe core-collapse event, 
because elliptical galaxies generally contain only low-mass, 
old population stars.
Recently, Hakobyan et al.$^{29}$ reexamined the morphology of the 
host galaxies of 22 core-collapse SNe (i.e., type II/Ibc) 
which had been previously classified as Elliptical or S0 
galaxies.  
They concluded that 19 cases were simply misclassifications of 
the host galaxy type.  
NGC 4589 remains a genuine E2 galaxy.

However, from the literature search related to NGC 4589, 
Hakobyan et al. pointed out that there is a 
Low Ionization Nuclear Emission-line Region 
(LINER) activity (Type ``L2''$^{38}$), 
and suggested the host to be a merger remnant.  
There is also an evidence for unusual distribution of 
interstellar dust from HST and AKARI studies$^{39,40}$, 
being consistent with the merger scenario.
Thus, the appearance of the SN Ib in this particular 
early-type galaxy may not conflict with the general 
scenario of stellar evolution and explosion.

According to a population synthesis model for the integrated 
light from the host galaxy$^{30}$, it has been 
suggested that about 90\% of the host flux is contributed by 
an old population with life times longer than $10^{10}$ years 
(i.e., $M_{\rm ms} \leq 1 M_{\odot}$), whereas a relatively 
young population with life times $\sim 10^7 - 10^8$ years 
(i.e., $M_{\rm ms} \leq 10 M_{\odot}$) contributes to the 
remaining $\sim 10$\%.  
Thus, it is likely that SN Ib 2005cz is the end product of 
one of these young stars, which were produced by the 
galaxy merger about $\sim 10^{8}$ years ago.

\section{Progenitors of faint, Ca-rich supernovae}

In addition to SN 2005cz, there are examples of
a faint, hydrogen deficient, possibly core-collapse SN 
which shows the large Ca/O line ratios, e.g., 
SNe 1997D, 2005E, 2005cs, 2005hk, and 2008ha$^{11,12,10,1}$.
As discussed below, however, they have very different 
observational features except for the late-time Ca/O ratio.
Thus it may not be the case that the origin of the large
Ca/O ratio and the progenitors are the same for all
these SNe.

SN 2005hk belongs to a subclass of peculiar SNe Ia,
SN 2002cx-like class$^{9,41}$, characterized by low luminosities 
and low expansion velocities ($\sim 5,000$ km s$^{-1}$).
The early-phase spectra of SN 2005hk are dominated by 
low-velocity permitted lines of Fe, without any resemblance 
to SN 2005cz whose velocity and spectral features are
those of typical SNe Ib (Supplementary Fig. 2).
Thus, it is unlikely that the progenitor of SN 2005hk
is the same as SN 2005cz.

The early spectra of SN 2008ha are similar to
SN 2005hk, but show even much lower velocities ($\sim 2,000$
km s$^{-1}$ for most lines, and $\sim 5,000$ km s$^{-1}$ 
for Ca II IR)$^{10,12}$. No clear detection of He is reported.
This is totally different from the early spectrum 
of SN 2005cz that shows strong He lines and the expansion 
velocity of $\sim 10,000$ km s$^{-1}$ (Supplementary Fig. 1).
The low expansion velocity of SN 2008ha is consistent with 
the fall-back SN model with a massive, black-hole-forming
progenitor rather than the less massive progenitor model$^{44}$.
It has also been claimed to be an explosion of a white dwarf
like other SNe Ia based on detection of silicon and sulfur
features in the early-phase$^{45}$. 
Thus, the origin of SN 2008ha is still controversial.

Adding to this, we should note that the reported 
``late-time'' spectra of SNe 2005hk and 2008ha 
are not fully nebular (Fig. 2), which is in contrast to 
the case of SN 2005cz.
Thus, it is possible that the O/Ca ratio is affected
by the attenuation within the ejecta. This would imply 
that the weak (or absence of) [O I] in these SNe would not 
necessarily indicate the small O-layer.
This effect may also appear in SN 2005E to some extent, given its
relatively young age ($\sim 2$ months) of the reported 
late-time spectrum$^{1}$.
Our late-time spectrum of SN 2005cz seems genuinely nebular,
and thus the large Ca/O ratio is more clearly the case than in
other examples (except for SNe II; see ref. 2 and below).

For SNe 2005hk and 2008ha, it is also not clear 
whether the strong Ca lines in the late-phases are emitted 
from the newly-synthesized materials.
The velocities of the ``nebular''
lines are similar to those in the early-phase spectra$^{12}$. 
In contrast, the velocity of the nebular Ca lines in 
SN 2005cz ($\sim 1,500$ km s$^{-1}$) is much lower than 
that in the early-phase (Supplementary Fig. 1), suggesting that 
the Ca lines are emitted from the innermost region of the ejecta 
where the newly-synthesized Ca dominates the emission.
  
SNe 1997D and 2005cs are both faint SNe II with 
the slow expansion velocities $^{46,47}$.
Although they do show the large Ca/O ratio in the nebular spectra, 
the comparison with SN Ib 2005cz should be done carefully. 
SNe II generally show the Ca/O line ratio being larger than SNe Ib/c, 
since Ca in the H-rich envelope can also contribute to [Ca II] and 
Ca II IR triplet (e.g., ref. 14).
A low mass progenitor is favoured for SN 2005cs, while the 
progenitor of SN 1997D is still controversial. 
These progenitors may or may not be consistent with 
the Ca/O line ratio in the nebular phase.
Further study including the emission from the H-rich
envelope is necessary to use the Ca/O line ratio 
as an indicator of the progenitor mass for SNe II. 

Detailed composition structure should be the key to 
the understanding of the progenitor and explosion 
mechanism of SN 2005cz. 
Deriving the detailed abundance from the nebular spectrum, 
however, is highly model dependent (e.g., see the 
above discussion for SNe II). 
Unfortunately, there is no strong Si or S line 
in optical wavelengths in the nebular phase, 
which could in principal be used to discriminate 
different scenarios.

The existence of 2005cz-like SNe, which have ejected 
material with little O and a relatively large amount 
of Ca, may have important implications to the chemical
evolution of galaxies.  In our Galaxy, a very Ca-rich,
extremely metal-poor (EMP) halo star has recently been
discovered$^{48}$.  Such an EMP star may have formed
from the debris of 2005cz-like SNe.  It would also be
interesting to note that some dwarf galaxies contain
EMP stars whose abundance ratios between the alpha-elements
and Fe are much smaller than the halo stars$^{49,50}$.
The oxygen-poor 2005cz-like SNe might be related to the 
formation of such EMP stars.

Current theoretical models still have lots of uncertainties
and further observational constraints are necessary to fully 
understand the final stage of the evolution of stripped
stars of different masses (See also Supplementary Fig. 3).
The evolutionary scenario of 10--12 $M_{\odot}$ we propose
in this paper (paragraph 5--6) is indeed similar to 
those have applied for ordinary SNe Ibc from more massive 
than 12 $M_{\odot}$.
However, it is a new theoretical argument that 10--12 
$M_{\odot}$ low mass models can have distinct properties of 
low $M(^{56}\rm Ni)$ production, low explosion energy, 
and the large Ca/O (see ref. 2 for the similar conclusions
from observations of SNe II).
These are quite different from more 
massive models, and consistent with the new 
observation of SN 2005cz.
Also, our suggestion to connect the explosions in the 
ONeMg white dwarfs with SNe Ib is quite new (see also ref. 1).
We also note that our discovery of SN 2005cz and the faint
nature of the 10--12 $M_{\odot}$ binary SN may solve 
the puzzle why SNe Ib from 10--12 $M_{\odot}$ binary stars 
have not been observed.  Even though such low mass stars 
should be more abundant in the Universe than 
more massive stars (e.g., the progenitor of SN 1994I),
they may simply be missed because of faintness.

\clearpage
\section{Supplementary Figure 1}

 \begin{center}
  \includegraphics[width=0.8\textwidth]{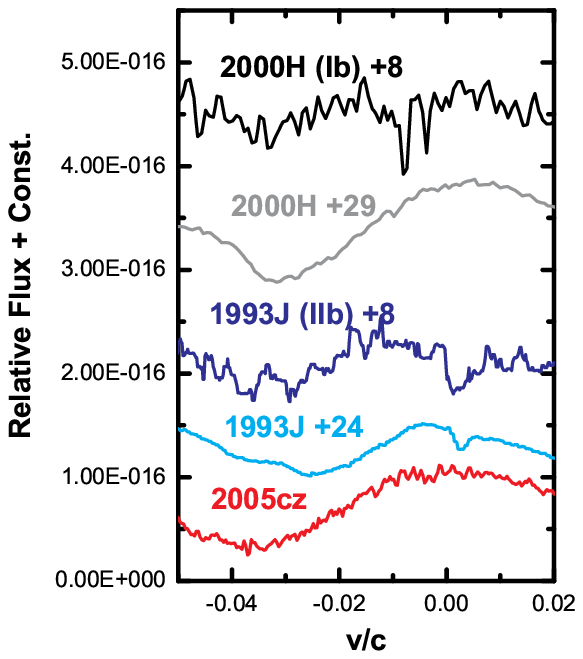}
 \end{center}

 A close-up plot of the He I $\lambda$5876 line in the early-phase
 spectrum of SN 2005cz and some SNe for comparison. 
 The horizontal axis denotes the line velocity normalized
 by the speed of light.
 The blueshift of the absorption component reaches $0.3-0.4c$
 for SN 2005cz.

\clearpage
\section{Supplementary Figure 2}

 \begin{center}
  \includegraphics[width=0.8\textwidth]{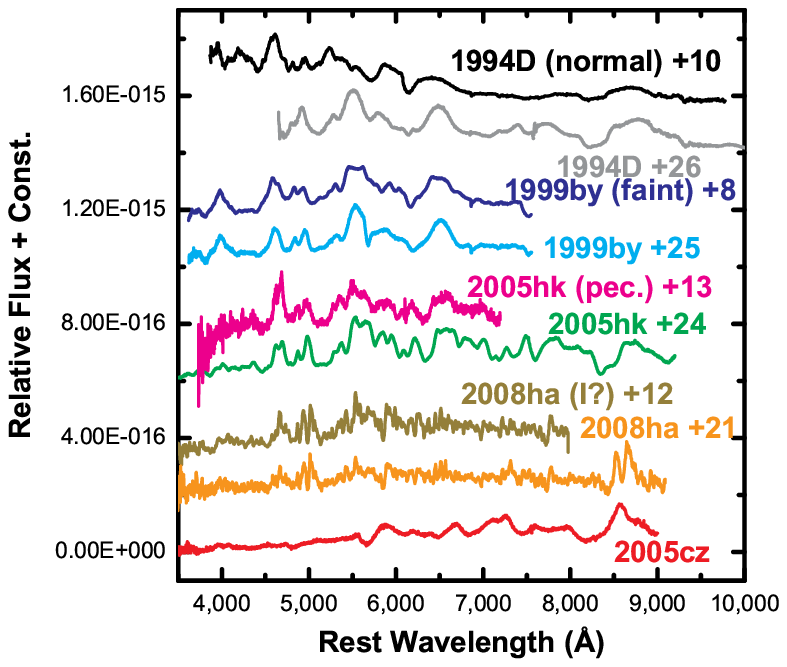}
 \end{center}

 Early-time spectrum of SN 2005cz in comparison with SNe Ia.
 From top to bottom, we show the normal SN Ia 1994D at $t=+10$
 and $+26$ days (ref. 42), the faint SN Ia 1999by at $t=+8$
 and $t=+25$ days (ref. 43), the peculiar SN Ia 2005hk at $t=+13$
 and $t=+24$ days (ref. 9), and the peculiar SN I? 2008ha at $t=+12$
 and $t=+21$ days (ref. 10); none of them is similar to 
 SN 2005cz on 2005 July 28 (presumably at $t=+26$ days).

\clearpage
\section{Supplementary Figure 3}

 \begin{center}
  \includegraphics[width=0.8\textwidth]{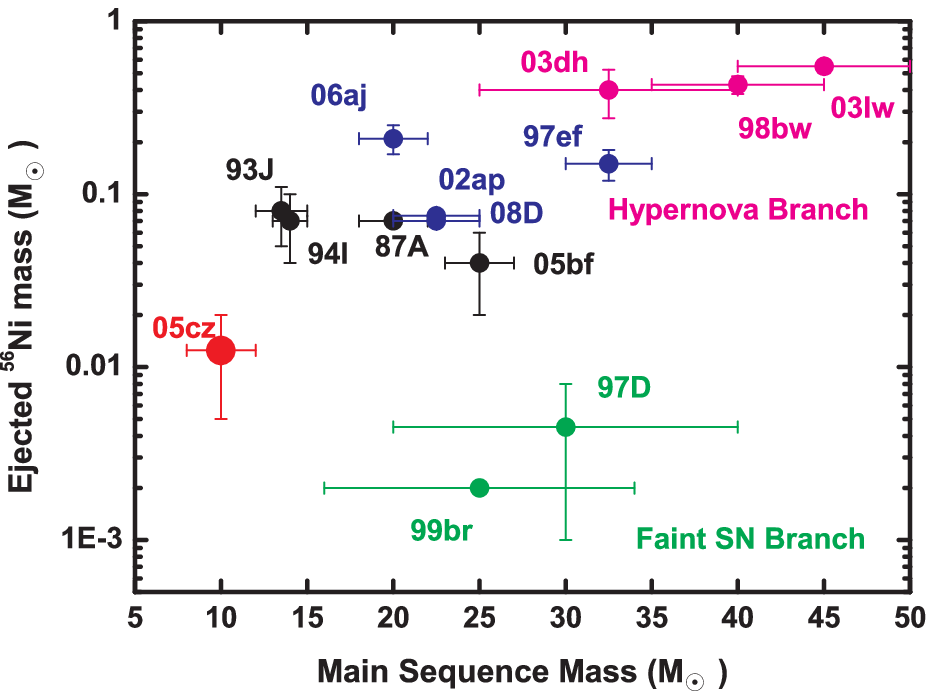}
 \end{center}

 The relation between progenitor mass and synthesized $^{56}$Ni 
 mass$^{7,16,23,51-68}$.
 The three SNe shown by magenta symbols at the upper right of the panel
 are associated with $\gamma$-ray bursts.
 These SNe and SN 1997ef are called hypernovae, with the definition
 that the kinetic energy of the explosion exceeds $10^{52}$ erg.
 SN 2005cz locates roughly at the bottom of the 
 sequence from hypernova to normal SNe.

\clearpage
\noindent
31. Puckett, T. \& Sehgal, A., Supernovae 2005cx, 2005cy, 2005cz,
	{\it IAU Circ.} {\bf 8569} (2005).\\
\noindent
32. Kashikawa, N. et al., FOCAS: The Faint Object Camera and
	Spectrograph for the Subaru Telescope, {\it Publ. Astron. Soc. Japan} 
	{\bf 54}, 819-832 (2002).\\
\noindent
33. Tomita, H. et al., The optical/near-infrared light curves
	of SN 2002ap for the first 1.5 years after discovery, {\it Astrophys.
	J.} {\bf 644}, 400-408 (2006).\\
\noindent
34. Oke, J. B. et al., The Keck Low-Resolution Imaging
	Spectrometer, {\it Publ. Astron. Soc. Pacif.} {\bf 107},
	375-385 (1995).\\
\noindent
35. Schlegel, D. J., Finkbeiner, D. P. \& Davis, M.,
	Maps of Dust Infrared Emission for Use in Estimation of Reddening
	and Cosmic Microwave Background Radiation Foregrounds,
	{\it Astrophys. J.} {\bf 500}, 525-553 (1998).\\
\noindent
36. Turatto, M., Benetti, S., Cappellaro, E., Variety in Supernova, in 
	{\it From Twilight to Highlight: The Physics of Supernovae}
	(eds. Hillebrandt, W. \& Leibundgut, B.) 200-209 (Springer,
	Berlin, 2003).\\
\noindent
37. Goudfrooij, P. 1999, The nature of Ionized Gas in Giant Elliptical Galaxies,
    in {\it Star Formation in Early-Type
	Galaxies} (eds. Cepa, J. \& Carral, P.), ASP Conf. Ser. 163,
	55-71 (ASP, San Francisco, 1999).\\
\noindent
38. Ho, L. C., Filippenko, A. V. \& Sargent, W. L. W.,
	A search for ``dwarf'' Seyfert nuclei. III. spectroscopic
	parameters and properties of the host galaxies,
	{\it Astrophys. J. Suppl.} {\bf 112}, 315-390 (1997).\\
\noindent
39. Tran, H. D. et al., Dusty nuclear disks and filaments in early-type 
	galaxies, {\it Astrophys. J.} {\bf 121}, 2928-2942 (2001).\\
\noindent
40. Kaneda, H., Suzuki, T., Onaka, T., Okada, Y., \& Sakon, I., 
        Spatial Distributions of Dust and Polycyclic Aromatic
	Hydrocarbons in the Nearby Elliptical Galaxy NGC 4589
	Observed with AKARI, {\it Publ. Astron. Soc. Japan} 
	{\bf 60}, S467-S475 (2008).\\
\noindent
41. Sahu, D. K. et al., The evolution of the peculiar type Ia Supernova
        SN 2005hk over 400 days, {\it Astrophys. J.} {\bf 680}, 580-592 (2008).\\
\noindent
42. Patat, F. et al., The type Ia supernova 1994D in NGC 4526: 
	the early phases, {\it Mon. Not. R. Astron. Soc.} {\bf 278},
	111-124 (1996).\\
\noindent
43. Garnavich, P. M. et al., The luminosity of SN 1999by in 
	NGC 2841 and the nature of ``peculiar'' type Ia supernovae,
	{\it Astrophys. J.} {\bf 613}, 1120-1132 (2004).\\
\noindent
44. Moriya, T. et al., Faint supernovae with fall-back, 
    a poster paper presented at KITP conference: 
    {\it Stellar Death and Supernovae} (August 17-21, 2009),
    http://online.kitp.ucsb.edu/online/sdeath\_c09/moriya.\\
\noindent
45. Foley, R. J. et al., Early- and late-time observations of 
    SN 2008ha: Additional constraints for the progenitor and 
    explosion, {\it Astrophys. J.} {\bf 708}, L61-L65 (2010).\\
\noindent
46. Turatto, M. et al., The peculiar type II supernova 1997D:
        A case for a very low $^{56}$Ni mass,
	{\it Astrophys. J.} {\bf 498}, L129-L133 (1998).\\
\noindent
47. Pastorello, A. et al., SN 2005cs in M51 -- II. Complete evolution
        in the optical and the near-infrared,
	{\it Mon. Not. R. Astron. Soc.} {\bf 394}, 2266-2282 (2009).\\
\noindent
48. Lai, D. K. et al., A Unique Star in the Outer Halo of the 
	Milky Way, {\it Astrophys. J.} {\bf 697}, L63-L67 (2009).\\
\noindent
49. Tolstoy, E., Hill, V., \& Tossi, M., Star formation 
        histories, abundances and kinematics of dwarf galaxies in the 
        Local group, {\it Ann. Rev. Astron. Astrophys} {\bf 47}, 371-425
	(2009).\\
\noindent
50. Aoki, W. et al., Chemical composition of extremely 
        metal-poor stars in the sextans dwarf spheroidal galaxy, {\it Astron.
        Astrophys.} {\bf 502}, 569-578 (2009).\\
\noindent
51. Shigeyama, T., \& Nomoto, K., Theoretical light curve of SN 1987A and 
        mixing of hydrogen and nickel in the ejecta, 
    {\it Astrophys. J.} {\bf 360}, 242-256 (1990).\\
\noindent
52. Blinnikov, S. et al., radiation hydrodynamics of SN 1987A. 
    I. Global Analysis of the Light Curve for the First 4 Months, 
    {\it Astrophys. J.} {\bf 532}, 1132-1149 (2000).\\
\noindent
53. Turatto, M. et al., The peculiar type II supernova 1997D: 
    A case for a very low $^{56}$Ni Mass, 
    {\it Astrophys. J.} {\bf 498}, L129-L133 (1998).\\
\noindent
54. Iwamoto, K. et al., The peculiar type Ic supernova 1997ef: 
    Another hypernova, {\it Astrophys. J.} {\bf 534}, 660-669 (2000).\\
\noindent
55. Mazzali, P. A., Iwamoto, K., \& Nomoto, K., A spectroscopic 
    analysis of the energetic type Ic hypernova SN 1997ef, 
    {\it Astrophys. J.} {\bf 545}, 407-419 (2000).\\
\noindent
56. Nakamura, T. et al., Light curve and spectral models 
    for the hypernova SN 1998bw associated with GRB 980425, 
    {\it Astrophys. J.} {\bf 550}, 991-999 (2001).\\
\noindent
57. Iwamoto, K. et al., A hypernova model for the supernova 
    associated with the $\gamma$-ray burst of 25 April 1998, 
    {\it Nature} {\bf 395}, 672-674 (1998).\\
\noindent
58. Zampieri, L. et al., Peculiar, low-luminosity type II 
    supernovae: low-energy explosions in massive progenitors?, 
    {\it Mon. Not. R. Astron. Soc.} {\bf 338}, 711-716 (2003).\\
\noindent
59. Mazzali, P. A. et al., The type Ic hypernova SN 2002ap, 
    {\it Astrophys. J.} {\bf 572}, L61-L65 (2002).\\
\noindent
60. Mazzali, P. A. et al., The type Ic Hypernova 
    SN 2003dh/GRB 030329, {\it Astrophys. J.} {\bf 599}, L95-L98 (2003).\\
\noindent
61. Deng, J. et al., On the light curve and spectrum of 
    SN 2003dh separated from the optical afterglow of GRB 030329, 
    {\it Astrophys. J.} {\bf 624}, 898-905 (2005).\\
\noindent
62. Mazzali, P. A. et al., Models for the type Ic hypernova 
    SN 2003lw associated with GRB 031203, 
    {\it Astrophys. J.} {\bf 645}, 1323-1330 (2006).\\
\noindent
63. Tominaga, N. et al., The unique type Ib supernova 2005bf: 
    A WN star explosion model for peculiar light curves 
    and spectra, {\it Astrophys. J.} {\bf 633}, L97-L100 (2005).\\
\noindent
64. Maeda, K. et al., The unique type Ib supernova 2005bf
     at nebular phases: A possible birth event of a strongly magnetized
     neutron star, {\it Astrophys. J.} {\bf 666} 1069-1082 (2007).\\
\noindent
65. Mazzali et al., A neutron-star-driven X-ray flash associated 
    with supernova SN 2006aj, {\it Nature} {\bf 442}, 1018-1020 (2006).\\
\noindent
66. Mazzali, P. A. et al., The metamorphosis of supernova SN 2008D/XRF 080109:
    A link between supernova and GRBs/hypernovae, {\it Science}
    {\bf 5893}, 1185-1188 (2008).\\
\noindent
67. Tanaka, M. et al., Type Ib supernova 2008D associated 
    with the luminous X-ray transient 080109: An 
    energetic explosion of a massive helium star, 
    {\it Astrophys. J.} {\bf 692}, 1131-1142 (2008).\\
\noindent
68. Nomoto, K., Maeda, K., Umeda, H., Ohkubo, T., Deng, J., \& Mazzali, P., 
    Hypernovae and their nucleosynthesis, in {\it A massive star
    odyssey, from main sequence to supernova} (eds. van der Hucht, K.A.,
    Herrero, A., \& Esteban, C.) IAU Symp. 212, 395-403 
    (San Francisco: ASP, 2003), astro-ph/0209064.\\


\begin{thebibliography}{11}
\bibitem{1} Perets, H. B. et al., A new type of stellar explosion,
        {\it Nature}, submitted, arXiv:0906.2003 (2009).
\bibitem{2} Smartt, S. J., Progenitors of core-collapse supernovae,
        {\it Ann. Rev. Astron. Astrophys.}, {\bf 47}, 63-106 (2009).
\bibitem{3} Zhang, Y., Gu, Q.-S. \& Ho, L. C., Stellar and dust
	properties of local elliptical galaxies: clues to the onset of
	nuclear activity, {\it Astron. Astrophys.} {\bf 487}, 177-183 (2008).
\bibitem{4} Leonard, D. C., Supernova 2005cz in NGC 4589, {\it IAU Circ.}
	{\bf 8579} (2005).
\bibitem{5} Branch, D. et al., Direct Analysis of Spectra of Type Ib
	Supernovae, {\it Astrophys. J.} {\bf 566}, 1005-1017 (2002). 
\bibitem{6} Barbon, R. et al., SN 1993J in M81: One year of observations 
	at Asiago, {\t Astron. Astrophys. Suppl.} {\bf 110}, 513-519 (1995).
\bibitem{7} Filippenko, A. V. et al., The Type Ic supernova 1994I in
	M51: Detection of helium and spectral evolution, {\it Astrophys. J.}
	{\bf 450}, L11-L15 (1995).
\bibitem{8} Sauer, D. N. et al., The properties of the `standard' Type
	Ic supernova 1994I from spectral models,
	{\it Mon. Not. R. Astron. Soc.} {\bf 369}, 1939-1948 (2006).
\bibitem{9} Maeda, K. et al., Asphericity in Supernova Explosions from
	Late-Time Spectroscopy, {\it Science} {\bf 319}, 1220-1223 (2008).
\bibitem{10} Phillips, M. M. et al., The peculiar SN 2005hk: Do some 
	type Ia supernovae explode as deflagrations?,
	{\it Publ. Astron. Soc. Pacif.} {\bf 119}, 360-387 (2007).
\bibitem{11} Valenti, S. et al., A low energy core-collapse supernova
	without a hydrogen envelope, {\it Nature} {\bf 459}, 674-677 (2009).
\bibitem{12} Filippenko, A. V. et al., Supernovae 2001co, 2003H, 2003dg, 
	and 2003dr, {\it IAU Circ.} {\bf 8159}, 2 (2003).
\bibitem{13} Foley, R. J. et al., SN 2008ha: An extremely low luminosity
        and exceptionally low energy supernova, 
	{\it Astron. J.} {\bf 138}, 376-391 (2009).
\bibitem{14} Nomoto, K., Tominaga, N., Umeda, H., Kobayashi, C. \&
	Maeda, K., Nucleosynthesis yields of core-collapse and hypernovae,
	and galactic chemical evolution, {\it Nucl. Phys. A}, 
	{\bf 777}, 424-458 (2006).
\bibitem{15} Fransson, C. \& Chevalier, R. A., Late emission from
	supernovae: A window on stellar nucleosynthesis, {\it Astrophys. J.}
	{\bf 343}, 323-342 (1989).
\bibitem{16} Maeda, K. et al., SN 2006aj Associated with XRF 060218 at
	Late Phases: Nucleosynthesis signature of a neutron-driven 
	explosion, {\it Astrophys. J.} {\bf 658}, L5-L8 (2007).
\bibitem{17} Shigeyama, T. et al., Theoretical light curves of type IIb
	Supernova 1993J, {\it Astrophys. J.} {\bf 420}, 341-347 (1994).
\bibitem{18} Nomoto, K. et al., A carbon-oxygen star as progenitor of
	the type-Ic supernova 1994I, {\it Nature} {\bf 371}, 227-229 (1994).
\bibitem{19} Mazzali, P. A., Nomoto, K., Patat, F. \& Maeda, K.,
    The nebular spectra of the hypernova SN 1998bw and evidence for
    asymmetry, {\it Astrophys. J.} {\bf 559}, 1047-1053 (2001).
\bibitem{20} Mazzali, P. A. et al., The aspherical properties of the
     energetic type Ic SN 2002ap as inferred from its nebular spectra,
     {\it Astrophys. J.} {\bf 670}, 592-599 (2007).
\bibitem{21} Richmond, M. W. et al., $UBVRI$ photometry of the type Ic 
	SN 1994I in M51, {\it Astrophys. J.} {\bf 111}, 327-339 (1996).
\bibitem{22} Stritzinger, M. et al., The He-rich Core-collapse Supernova
	2007Y: Observation from X-ray to Radio Wavelengths,
	{\it Astrophys. J.} {\bf 696}, 713-728 (2009).
\bibitem{23} Botticella, M. T. et al., SN 2008S: an electron capture SN
	from a super-AGB progenitor?, {\it Mon. Not. R. Astron. Soc.},
	accepted, arXiv: 0903.1286 (2009).
\bibitem{24} Iwamoto, K. et al., Theoretical light curves for the Type
	Ic Supernova SN 1994I, {\it Astrophys. J.} {\bf 437}, L115-L118 (1994).
\bibitem{25} Patat, F. et al., The Metamorphosis of SN 1998bw,
	{\it Astrophys. J.} {\bf 555}, 900-917 (2001).
\bibitem{26} Yoshii, Y. et al., The optical/near-infrared light curves 
	of SN 2002ap for the first 140 days after discovery,
	{\it Astrophys. J.} {\bf 592}, 467-474 (2003).
\bibitem{27} Tanaka, M. et al., Nebular phase observations of the Type
	Ib Supernova 2008D/X-ray transient 080109: Side-viewed bipolar
	explosion, {\it Astrophys J.} {\bf 700}, 1680-1685 (2009).
\bibitem{28} Nomoto, K. \& Hashimoto, M., Presupernova evolution of
	massive stars, {\it Phys. Rep.} {\bf 163}, 13-36 (1988).
\bibitem{29} Nomoto, K., Evolution of 8-10 solar mass stars toward
	electron capture supernovae. I - Formation of
	electron-degenerate O + Ne + Mg cores, 
	{\it Astrophys. J.} {\bf 277}, 791-805 (1984).
\bibitem{30} Hakobyan, A. A. et al. 2008, Early-type galaxies with core
	collapse supernovae, {\it Astron. Astrophys.} {\bf 488}, 
	523-531 (2008).
\end{thebibliography}
\end{document}